# Theoretical Study on MR-TADF Materials Based on CzBN

**Jinpu Bai [a], Lu Shen [c], Aynur Matyusup [a], Aimin Ren [b], and Jingfu Guo [a,*]**

[a]College of physics, Northeast Normal University, Changchun 130024;

[b] Institute of theoretical chemistry, College of Chemistry, Jilin University, Changchun 130023

[c] Department of basic science, Jilin jianzhu university，Changchun，130118

*Corresponding author：Guo Jing-fu, E-mail address: guojf217@nenu.edu.cn.

†equally contributed to this work

**Abstract** Multi-resonance thermally activated delayed fluorescence (MR-TADF) materials have garnered significant research interest owing to their remarkably narrow emission spectra with full width at half maximum (FWHM) below 40 nm, demonstrating substantial advantages over conventional donor-acceptor (D-A) type TADF materials in spectral purity. However, conventional N-B-N resonant framework materials are fundamentally constrained by their intrinsically low reverse intersystem crossing rates (kRISC < $10^3$ $s^{-1}$), presenting a persistent challenge for achieving high-efficiency TADF. This study proposes a triple collaborative design strategy based on CzBN to break through this limitation: (1) Enhance the separation of HOMO and LUMO by π-conjugation expansion and reduce $\Delta_{EST}$; (2) The O/S heteroatom was introduced to control the excited state charge transfer (CT) characteristics to further reduce $\Delta_{EST}$; (3) The spin orbit coupling (SOC) effect is enhanced by the synergistic effect of the extended π system and heteroatoms. Based on this, five new MR-TADF molecules (CzBN_N, CzBN_O, CzBN_S, CzBN_Cz, CzBN_2S) were designed and studied. Among them, the $k_{RISC}$ of CzBN_S reached $3.48\times10^6$ $s^{-1}$ , which was two orders of magnitude higher than CzBN; At the same time, ΔEST < 0.1 eV and the FWHM maintained at 40 nm.

# 基于 CzBN 的有机多共振热活化延迟荧光材料的理论设计

白晋朴 [1†]，沈璐 [3†]，阿依努尔.麦提玉苏普 [1]，任爱民 [2]，郭景富 [1*]

[1] 东北师范大学物理学院，长春 130024

[2] 吉林大学化学学院理论化学研究所，长春 130023

[3] 吉林建筑大学 基础科学部 物理实验中心，长春 130118

*Corresponding author：Guo Jing-fu, E-mail address: guojf217@nenu.edu.cn.

†equally contributed to this work

## 摘要

多重共振热活化延迟荧光（MR-TADF）材料因其可实现窄发射光谱特性（半峰宽<40 nm）显著优于传统 D-A 型 TADF 材料而备受关注。然而，此类经典 N-B-N 共振骨架材料普遍存在反向系间窜越速率（$k_{RISC}$）偏低的瓶颈问题。本研究提出基于 CzBN 的三重协同设计策略以突破该限制：（1）通过 π 共轭平面拓展增强 HOMO、LUMO 轨道分离度，降低 $\Delta E_{ST}$；（2）引入 O/S 杂原子调控激发态电荷转移（CT）特性，进一步降低 $\Delta E_{ST}$；（3）利用扩展 π 体系与杂原子的协同作用增强旋轨耦合（SOC）效应。基于此，设计和研究了五种新型 MR-TADF 分子（CzBN_N、CzBN_O、CzBN_S、CzBN_Cz、CzBN_2S），其中 CzBN_S 的系间窜越速率 $k_{RISC}$ 达 $3.48\times10^6$ $s^{-1}$，较基础体系提升 2 个数量级；同时 $\Delta E_{ST}$ <0.1 eV 以及半峰宽维持 40 nm。

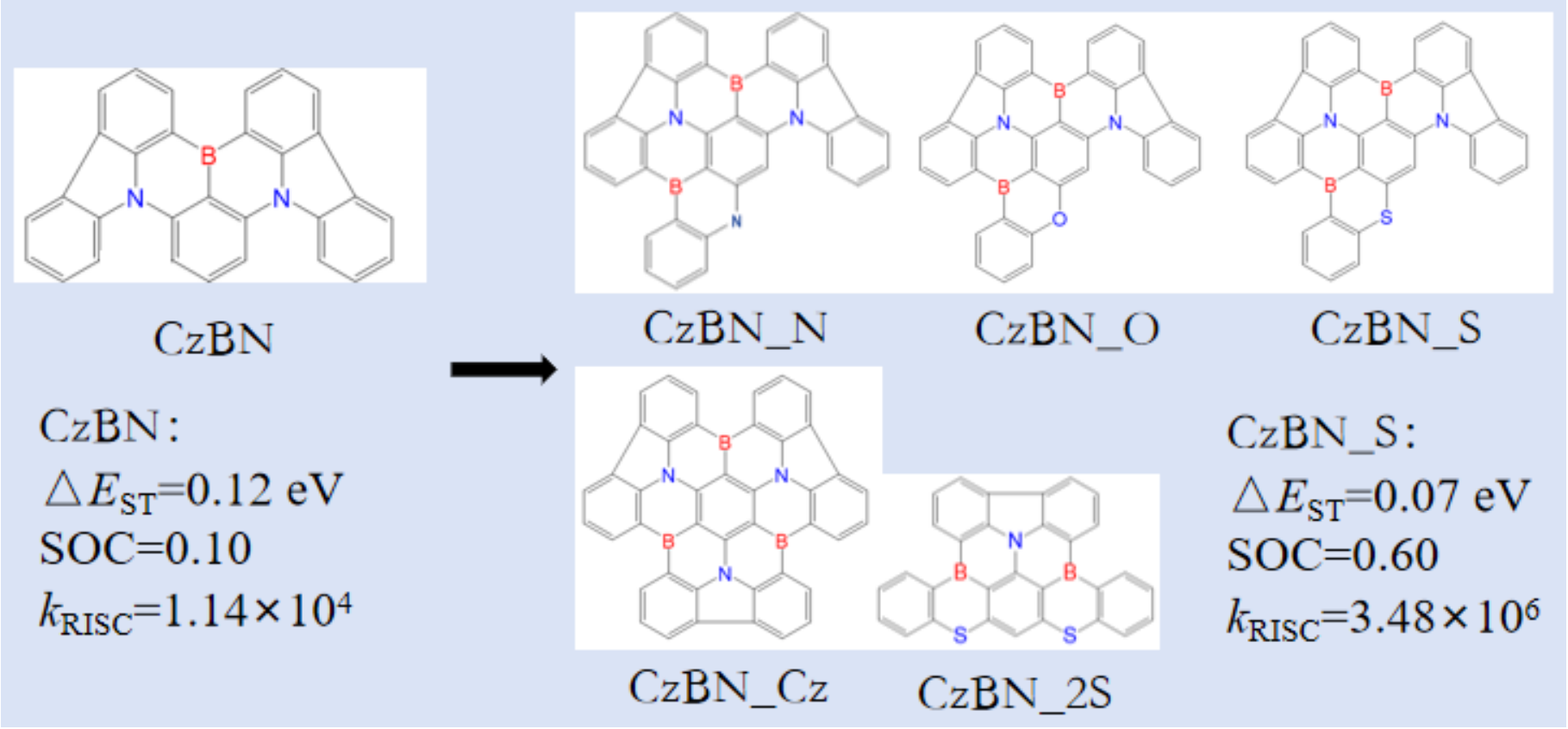

## 1 引言

有机发光二极管（OLED）的发展显著推动了固态照明和平板显示技术的革

新[1]。其中，磷光配合物的引入首次实现了三重态激子的有效利用[2]，而后续发展的热激活延迟荧光（TADF）材料进一步通过反向系间窜越（RISC）机制实现了单重态与三重态激子的全捕获，使 OLED 的内量子效率（IQE）在理论上可达100%发光。尤其值得注意的是，无金属 TADF 材料因其低成本、可调控的光物理特性及灵活的分子设计策略[3]，成为当前实现高效 OLED 器件的关键材料体系之一。

在热激活延迟荧光（TADF）材料的性能优化中，减小第一激发单重态（$S_1$）与三重态（$T_1$）之间的能隙（$\Delta E_{ST}$）是促进 $T_1 \rightarrow S_1$的反系间窜越（RISC）过程的核心策略之一，可显著提升激子利用率，从而优化 TADF 效率[4]。传统 TADF 材料设计通常采用高度扭曲的供体-受体（D-A）分子骨架，通过使最高占据分子轨道（HOMO）和最未占据分子轨道（LUMO）空间分离以降低电子交换能（exchange energy），进而实现 $\Delta E_{ST}$ 的精准调控[5]。此类材料被称为 D-A 型 TADF 材料。然而，D-A 型 TADF 材料存在固有局限性：其显著的斯托克斯位移（Stokes shift）和宽发射光谱（半峰宽通常达 70–100 nm）导致色纯度下降[6]。为实现基于 TADF 的 OLED 商业化，目前需依赖外源性光学调控技术（如滤光片或微腔结构）截断电致发光（EL）光谱的边缘区域以提升色纯度。但此类技术会引入额外的器件复杂度、能耗与制造成本[7]，严重制约 TADF-OLED 的市场竞争力。

2016 年，Hatakeyama 团队提出了一种基于刚性 π 共轭框架的多重共振（MR）分子设计策略[8]，该策略通过将硼（B）和氮（N）等电子受体/供体精准定位在多环芳烃（PAH）骨架中，构建了 MR-TADF 分子。此类分子的刚性平面结构有效抑制激发态下的分子弛豫，而其多重共振效应则实现了 HOMO、LUMO 轨道的空间分离，从而获得较小的 $\Delta E_{ST}$ 和窄带发射。但是，相较于 D-A-TADF 材料（$k_{RISC}$ 通常为 $10^5$–$10^7$ $s^{-1}$），多数 MR-TADF 材料的反向系间窜越（RISC）效率较低（$k_{RISC}$ 约 $10^4$ $s^{-1}$），限制了激子利用率及器件性能[9][10][11]。引入敏化剂可部分缓解三重态湮灭问题，但会伴随额外的能量损失[12][13]，制约了其商业化应用前景。

根据费米黄金法则，减小 $\Delta E_{ST}$ 和增强自旋-轨道耦合（SOC）均可有效提升 $k_{RISC}$。研究表明，在 MR-TADF 分子骨架中引入重原子（如 O、S、Se）可在保持前线轨道多重共振特性的同时增强 SOC 效应[14][15][16]。2023 年，黄教授团队[17]

基于 CzBN[18]和 BCzBN 结构设计了一系列新型 MR-TADF 分子（CzBN1-3），通过外围供电子基团的修饰实现了 $k_{RISC}$ 的提升。然而，该策略导致激子-声子耦合增大，增大了发射光谱的展宽（CzBN1 和 CzBN2 的 FWHM 从 23 nm 增至 75-78 nm），表明其性能仍需进一步优化。理论分析[17][19]表明，MR-TADF 分子的轨道跃迁主要表现为短程电荷转移（$^1$SRCT/$^3$SRCT），其 $^3$SRCT→$^1$SRCT 的 RISC 过程因自旋禁阻而呈现弱 SOC 特性。引入供电子基团后，体系产生新的长程电荷转移态（$^1$LRCT/$^3$LRCT），并使其成为低能态，此时 RISC 路径转变为 $^3$LRCT→$^1$SRCT/$^1$LRCT，该过程不仅规避了自旋禁阻，同时通过以下机制提升效率：（1）增强 SOC 效应；（2）降低有效 $\Delta E_{ST}$；（3）提供额外的系间窜越通道。

受上述研究启发，本研究基于 CzBN 分子骨架，通过 B-N-B 对位 π 共轭体系的平面扩展及 O/S 原子掺杂，设计了 CzBN_N、CzBN_O、CzBN_S、CzBN_Cz 和 CzBN_2S 等一系列新型 MR-TADF 分子。拟通过 π 共轭骨架扩展增强 HOMO-LUMO 轨道的离域分布，刚性结构维持 $S_0$-$S_1$ 态几何稳定性，高电负性 O/S 原子掺杂提升自旋轨道耦合（SOC），同时减小 $\Delta E_{ST}$，提升 $k_{RISC}$，进而实现高效率 MR-TADF 发光材料的设计。

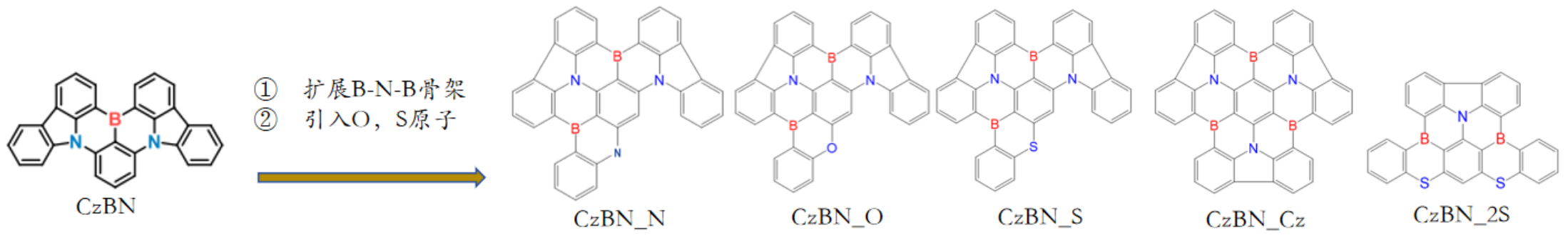

图 1　分子结构图

## 2　方法选择

本研究以 CzBN/BCzBN 为例，采用 Gaussian 16 程序[20]在 B3LYP/6-31+G(d,p) 水平上对 CzBN 和 BCzBN 分子进行基态和激发态的几何优化及电子结构计算，获得 HOMO-LUMO 电子密度分布；利用 MRCC 程序[21]基于 SCS-CC2/cc-pVDZ 方法计算 $\Delta E_{ST}$；通过 MOMAP 程序[22]中的 DUSHIN 模块计算了 $S_0$-$S_1$ 间重组能；采用 ORCA 程序[23]在 M062X/def2-TZVP 水平下计算旋轨耦合矩阵元数据理论计算结果与实验数据的如列表 1 所示，$\Delta E_{ST}$、SOC 和重组能等关键参数的计算误差均小于 5%，说明计算方法的选择是合理的。

表 1　理论计算系间窜越/反系间窜越相关的物理参数及实验测量值对比

| | $\lambda_{abs}$ (nm) | $\lambda_{em}$ (nm) | $\Delta E_{S1T1}$ | $\lambda_{S1\text{-}T1}$ (cm$^{-1}$) | $H_{S1\text{-}T1}(SOC)$(cm$^{-1}$) |
|---|---|---|---|---|---|

| | | | (eV) | $\lambda_{S1}$ | $\lambda_{T1}$ | $H_{S1}$ | $H_{T1}$ |
|---|---|---|---|---|---|---|---|
| $CzBN^{exp}$ | 458 | 477 | 0.14 | — | — | — | — |
| CzBN | 445.92 | 467.75 | 0.12 | 220 | 342 | 0.17 | 0.10 |
| $BCzBN^{exp}$ | 463 | 484 | 0.15 | — | — | — | — |
| BCzBN | 448.75 | 467.47 | 0.11 | 142 | 275 | 0.19 | 0.17 |

使用半经典的 Marcus 公式计算了系间窜越速率和反系间窜越速率[24]：

$$k_{(\mathrm{R})\mathrm{ISC}} = \frac{2\pi}{\hbar} < \mathrm{T}_1|H_{\mathrm{SOC}}|S_1 >^2 \frac{1}{\sqrt{4\pi\lambda k_{\mathrm{B}}T}} exp\left[-\frac{(\Delta E_{ST} \pm \lambda)^2}{4\lambda k_B T}\right] \quad (3.1)$$

使用爱因斯坦自发辐射方程计算荧光辐射跃迁速率[25]：

$$k_r^S = \frac{2\pi v^2 e^2}{\varepsilon_0 m c^3} f_{S_1 \to S_0} \quad (3.2)$$

相关速率计算结果，如表 2 所示：

表 2　理论计算的速率与实验测量值对比

| | $k_r$ ($s^{-1}$) | $k_{ISC}$ ($s^{-1}$) | $k_{RISC}$ ($s^{-1}$) |
|---|---|---|---|
| $CzBN^{exp}$ | $1.3\times10^8$ | $4.5\times10^7$ | $1.2\times10^4$ |
| CzBN | $1.3\times10^8$ | $1.1\times10^7$ | $1.1\times10^4$ |
| $BCzBN^{exp}$ | $1.4\times10^8$ | $7.2\times10^7$ | $1.4\times10^4$ |
| BCzBN | $1.5\times10^8$ | $1.2\times10^7$ | $1.4\times10^4$ |

由表 2 可知，计算的 $k_r$、$k_{ISC}$ 及 $k_{RISC}$ 均和实验测量值在同一个数量级，表明本文所采用的理论计算是适合 MR-TADF 分子的，可以准确的估算其分子性质和发光性质。

# 3 结果与讨论

## 3.1　基态前线分子轨道分布

在 B3LYP/6-31+G(d,p)水平优化的基态结构基础上，将分子的前线分子轨道 HOMO 和 LUMO 轨道的重叠程度及轨道分布图列于表 3 和图 2 中。

表 3　分子的 HOMO-LUMO 轨道重叠程度

| | CzBN | CzBN_N | CzBN_O | CzBN_S | CzBN_Cz | CzBN_2S |
|---|---|---|---|---|---|---|
| 重叠程度 (%) | 0.58 | 0.54 | 0.57 | 0.56 | 0.66 | 0.53 |

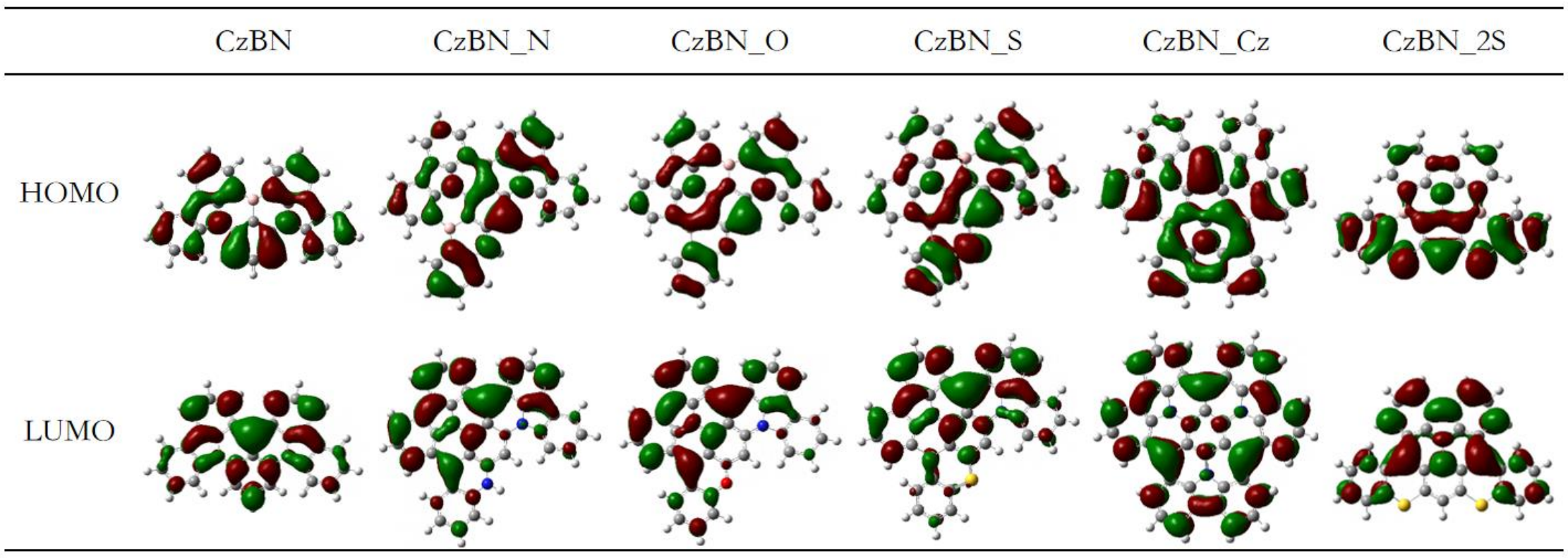

图 2　分子基态的前线分子轨道分布

如图 2 和表 3 所示，CzBN 分子呈现 HOMO 轨道（主要定域于 N 原子区域）与 LUMO 轨道（主要定域于 B 原子区域）的交替分布特征（轨道重叠度~58%），即多重共振效应的电子结构特征，以确保窄带发射特性；新设计分子在核心骨架区域完全保留了这一分布模式，H-L 轨道重叠度维持在 53~66%左右，核心电子分布特征表明其仍属于典型的 MR-TADF 材料体系。除 CzBN_2S（轨道重叠度 0.66）外，设计分子的 H-L 轨道重叠度均小于 0.58，HOMO-LUMO 轨道的空间分离度都优于 CzBN 分子，可能有利于降低电子交换能、减小 $\Delta E_{ST}$。

### 3.2　几何弛豫和电子-振动耦合分析

为了明确本文所设计分子的窄带发射特性，，我们对分子激发态结构的刚性及其电子振动耦合参数进行评估。首先在 B3LYP/6-31+G(d,p)/TD B3LYP/6-31+G(d,p)水平上对研究分子的基态 $S_0$ 和激发态 $S_1$ 进行了几何优化，其最低频率均为正值。在优化结构基础上，我们计算了设计分子 $S_0$ 和 $S_1$ 态间几何均方差 RMSD 值如表 4 所示（分别为 CzBN:0.079；CzBN_N:0.033；CzBN_O:0.065；CzBN_S:0.072；CzBN_Cz:0.011；CzBN_2S:0.098），结果表明，相比基态，其 $S_1$ 态的 RMSD 值均小于 0.1，除 CzBN_2S，设计分子的 $S_1$ 态的结构弛豫均小于实验分子 CzBN，表明这些分子 $S_1$ 态结构弛豫很小，分子结构有刚性较强。

表 4　分子 $S_0$ 和 $S_1$ 间几何均方差 RMSD 值

| | CzBN | CzBN_N | CzBN_O | CzBN_S | CzBN_Cz | CzBN_2S |
|---|---|---|---|---|---|---|
| $RMSD_{S0\text{-}S1}$ | 0.072 | 0.033 | 0.065 | 0.072 | 0.011 | 0.098 |

其次，我们用正则模分析法计算了激发态的重组能用以表征电子-振动耦合效应。如图 3 所示，CzBN 的 $S_1$ 态重组能为 475 $cm^{-1}$，对重组能贡献较大的振动模式位于 83 $cm^{-1}$（剪式振动）和 735 $cm^{-1}$（伸缩振动）频率处，重组能分别为 84 cm$^{-1}$ 和 56 cm$^{-1}$，占总重组能的 29%。结构修饰后，除 CzBN_Cz（重组能 431 cm$^{-1}$）外，其余设计分子（CzBN_N:535 cm$^{-1}$; CzBN_O:664 cm$^{-1}$; CzBN_S:613 cm$^{-1}$; CzBN_2S:764 cm$^{-1}$）重组能均有轻微增大。对 CzBN_N 的重组能贡献较大的振动模式位于 1412 cm$^{-1}$（伸缩振动）和 1560 cm$^{-1}$（剪式振动）频率处；对 CzBN_O 的重组能贡献较大的振动模式位于 438 cm$^{-1}$（剪式振动）和 1550 cm$^{-1}$（伸缩振动）频率处（低频处 C-O 键振动显著增加）；对 CzBN_S 的重组能贡献较大的振动模式位于 60 cm$^{-1}$（剪式振动）和 1525 cm$^{-1}$（伸缩振动）频率处，对 CzBN_Cz 的重组能贡献较大的振动模式位于 508 cm$^{-1}$（剪式振动）和 1147 cm$^{-1}$（伸缩振动）频率处，对 CzBN_2S 的重组能贡献较大的振动模式位于 115 cm$^{-1}$（剪式振动）和 1527 cm$^{-1}$（伸缩振动）频率处（低频处两 C-S 键振动显著增加），具体而言：（1）C-N 拓展（CzBN_N）使中心苯环伸缩振动频率提升至 1412 与 1560 cm$^{-1}$；（2）C-O/S 修饰（CzBN_O/S）分别在 1550/1525 cm-1 处产生新的拉伸振动；（3）双 S 原子引入（CzBN_2S）导致 C-S 键在 1527 cm-1 处产生显著拉伸振动。这些杂环或杂原子融入的 pi 结构扩展修饰增大了高频区振动的幅度（骨架伸缩振动），造成重组能的增大，这可能会对材料的窄带发射特性产生不利影响。除了 B-N-B 杂环引入的 CzBN_Cz 分子产生小于实验分子 CzBN 的重组能外，B-N 杂环（CzBN_N）、双 S 杂环（CzBN_2S）、C-S 杂环（CzBN_ S）和 C-O 杂环（CzBN_O）融入 CzBN 结构时电子-振动耦合效应依次增大，不利于发射光谱窄带化。

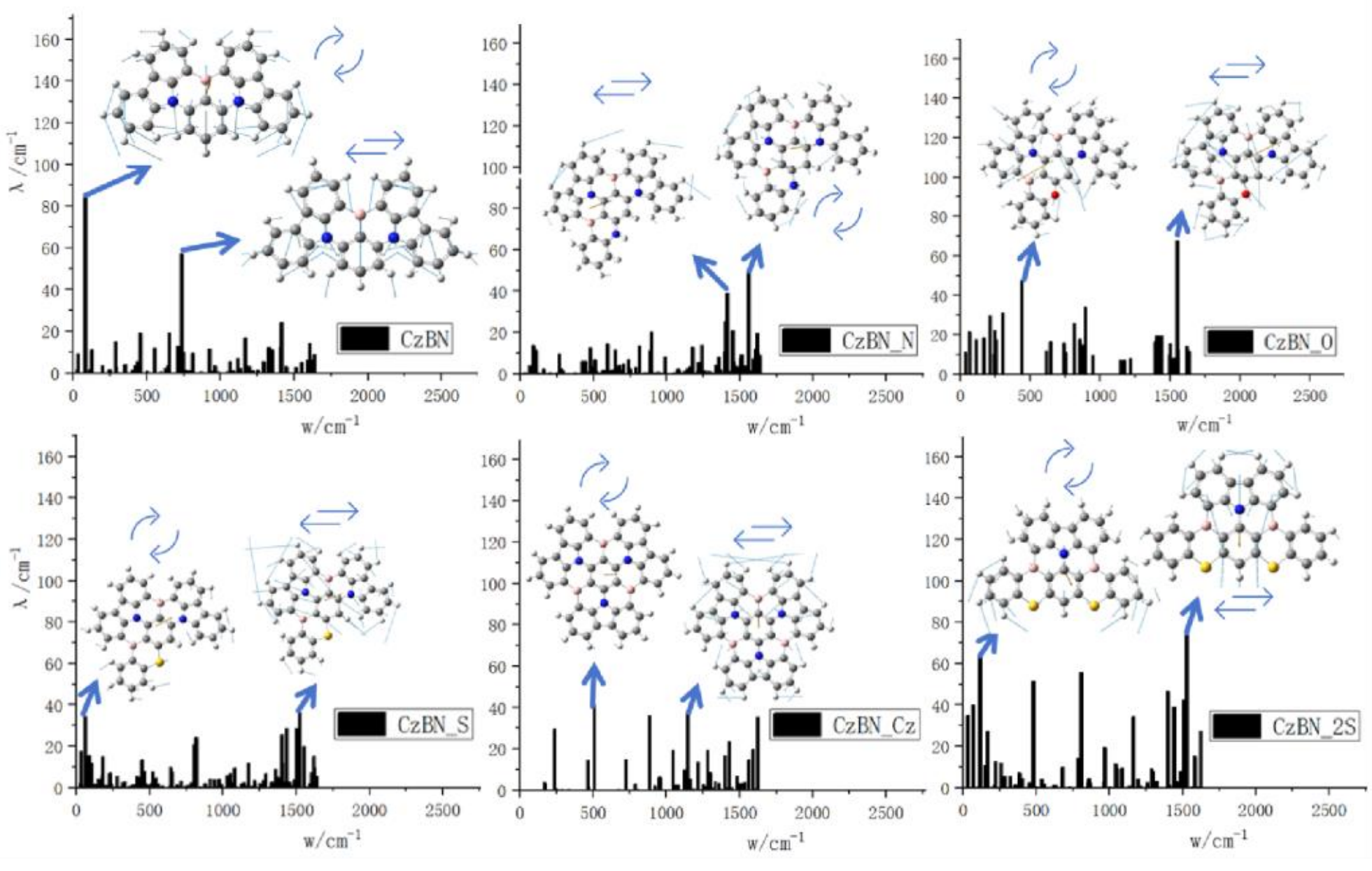

图 3　所研究分子 S$_1$ 态的振动分析

### 3.3 光谱分析

本文对研究分子的电子吸收和荧光发射光谱进行了计算，结果列于。

表-表 7 中。

表 5　分子的吸收和荧光发射波长，振子强度以及 HOMO-LUMO 能级

| | CzBN | CzBN_N | CzBN_O | CzBN_S | CzBN_Cz | CzBN_2S |
|---|---|---|---|---|---|---|
| $HOMO_{S0}$ (eV) | -5.55 | -5.51 | -5.71 | -5.66 | -5.92 | -5.71 |
| $LUMO_{S0}$ (eV) | -2.23 | -2.15 | -2.23 | -2.24 | -2.29 | -2.26 |
| HOMO-$LUMO_{S0}$(eV) | 3.32 | 3.36 | 3.48 | 3.42 | 3.63 | 3.45 |
| $\lambda_{abs}$ (nm) | 445.92 | 436.98 | 420.03 | 427.83 | 398.22 | 426.19 |
| $f_{abs}$ | 0.48 | 0.21 | 0.20 | 0.21 | 0.67 | 0.29 |
| $\lambda_{em}$ (nm) | 467.75 | 459.63 | 438.76 | 448.82 | 413.50 | 451.11 |
| $f_{em}$ | 0.43 | 0.28 | 0.40 | 0.37 | 0.39 | 0.36 |
| *Stokes shift* | 21.83 | 22.65 | 18.73 | 20.09 | 15.28 | 24.92 |

表 6　TD(B3LYP/6-31+G (d，p)计算下的各分子吸收光谱

| | Excited states | $\lambda_{abs}$ (nm) | $f_{abs}$ | Main configurations |
|---|---|---|---|---|
| CzBN | $S_1$ | 445.92 | 0.48 | H→L(99%) |
| CzBN_N | $S_1$ | 436.98 | 0.21 | H→L(98%) |
| CzBN_O | $S_1$ | 420.03 | 0.20 | H→L(98%) |
| CzBN_S | $S_1$ | 427.83 | 0.21 | H→L(97%) |
| CzBN_Cz | $S_1$ | 398.22 | 0.67 | H→L(98%) |
| CzBN_2S | $S_1$ | 426.19 | 0.29 | H→L(99%) |

表 7　TD(B3LYP/6-31+G（d，p）计算下的各分子发射光谱

| | Excited states | $\lambda_{em}$ (nm) | $f_{em}$ | Main configurations |
|---|---|---|---|---|
| CzBN | $S_1$ | 467.75 | 0.43 | H→L(99%) |
| | $S_2$ | 366.13 | 0.03 | H-3→L(11%)；H-1→L(81%)<br>H→L+1(5%) |
| CzBN_N | $S_1$ | 459.63 | 0.28 | H→L(99%) |
| | $S_2$ | 415.81 | 0.18 | H-1→L(12%)；H→L+1(86%) |
| CzBN_O | $S_1$ | 438.76 | 0.40 | H→L(98%) |
| | $S_2$ | 411.95 | 0.18 | H-1→L(4%)；H→L+1(86%) |
| CzBN_S | $S_1$ | 448.82 | 0.37 | H→L(98%) |
| | $S_2$ | 430.11 | 0.17 | H-1→L(5%)；H→L+1(93%) |
| CzBN_Cz | $S_1$ | 413.50 | 0.39 | H→L(98%) |
| CzBN_2S | $S_1$ | 451.11 | 0.36 | H→L(99%) |
| | $S_2$ | 425.99 | 0.08 | H-1→L(6%)<br>H→L+1(92%) |

相比于分子 CzBN，新设计分子的吸收、发射波长均发生蓝移。由于杂原子

介入的π结构扩展修饰使电子云呈现轻微局域分布，使的 HOMO 轨道能级略下降，造成分子 HOMO-LUMO 能隙差增加，使得光谱蓝移。由表 5 可知，新设计分子的 LUMO 能级未出现明显变化，HOMO 能级有不同程度降低，造成 HOMO-LUMO 能隙增大。CzBN_N 分子的 HOMO-LUMO 能隙变化较小，而伴有 O，S 原子加入的 CzBN_O，CzBN_S 和 CzBN_2S 以及 CzBN_Cz 能隙增加明显。除 CzBN_Cz 外，相比 CzBN 分子新设计分子的荧光发射强度 $f$ 均有不同程度的减小，不利于瞬时荧光发射。综合分析 $S_1$ 态的振动分析、振子强度以及 Stokes 位移，新设计分子可能会不利于发射谱展宽的进一步窄化。另外，设计分子的 Stokes 位移较小（<25nm)。CzBN_O、 CzBN_S、CzBN_Cz 的 Stokes 位移都比实验分子 CzBN（21nm）的小，但 CzBN_N 和 CzBN_2S 的 Stokes 位移则略大，表明 B-O、B-S、B-N-B 杂环融入 CzBN 结构时激发耗散能量小，有利于激子利用率提高，而 B-N 杂环和双 S 杂环介入 CzBN 结构激发耗散能量大，不利于提高激子利用率。

**3.4 光谱展宽计算**

实现窄带发射是 MR-TADF 分子结构优化的目标之一。对研究分子的发射谱及其半峰宽(FWHM)进行了理论预测。通过使用 ORCA 程序的 ESD 模块（TDDFT B3LYP/DEF2-SVP 水平下计算）计算所研究分子的振动分辨光谱，进而通过得到的光谱图测量了研究分子的 FWHM，如图 4 所示。

实验得到的原始分子 CzBN 的 FWHM 为 25 nm。通过理论预测得到的 CzBN，CzBN_N，CzBN_O，CzBN_S，CzBN_2S 的 FWHM 分别为 28 nm，29 nm，41 nm，43 nm，39 nm。表明 B-N 杂环（CzBN_N）、双 S 杂环（CzBN_2S）、B-O/S（CzBN_O/S）融入 CzBN 结构时发射电子-振动耦合效应依次增大，不利于发射光谱窄带化。这与电子-振动耦合分析结果一致。设计分子的 FWHM 与原始分子 CzBN 相比，有所增大，但是其半峰值仍然在窄带光谱范围内。

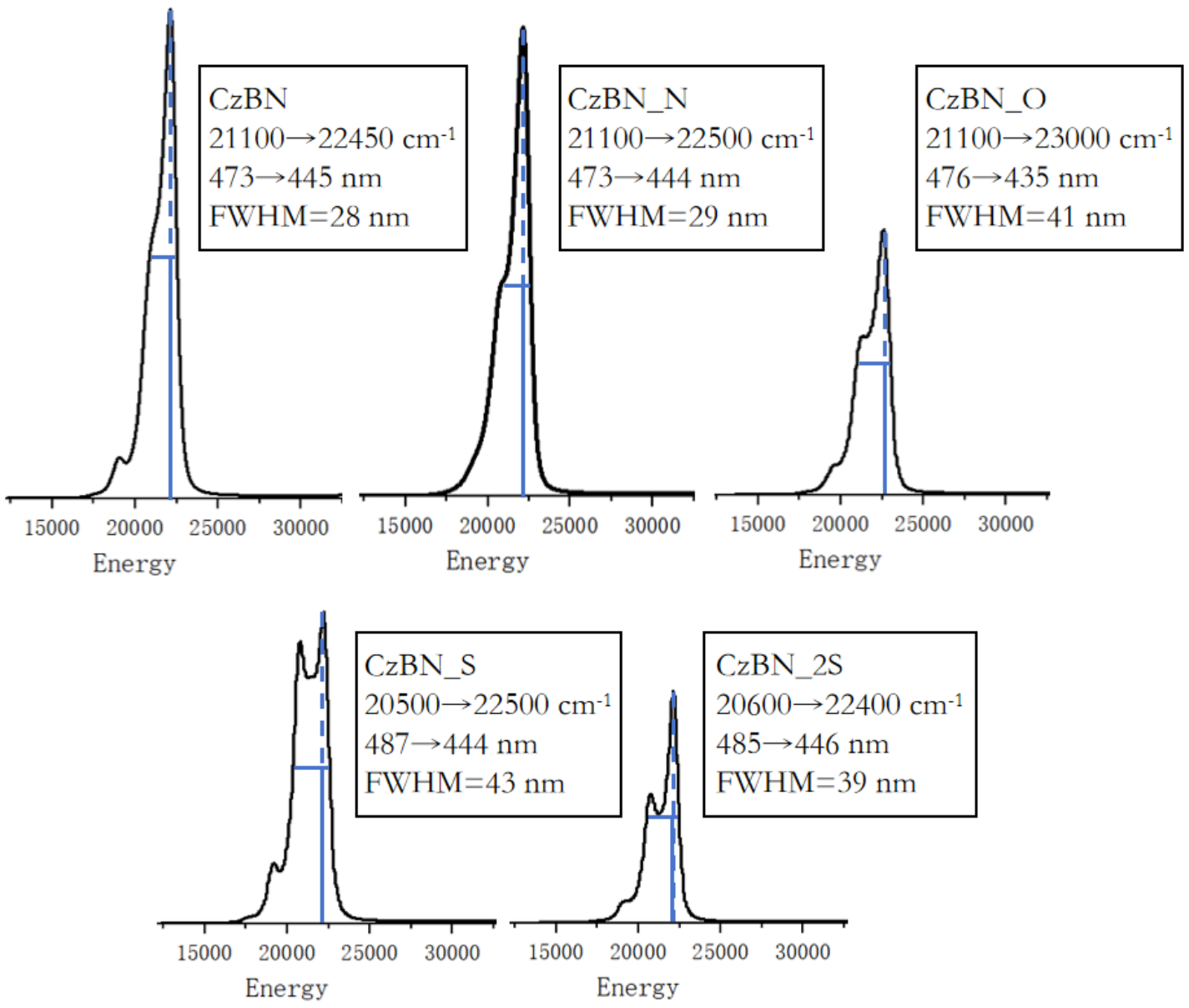

图 4　利用 ORCA 软件使用 B3LYP/DEF2-SVP 方法计算的振动分辨发射光谱及估算的 FWHM

### 3.5　单重和三重激发态前线分子轨道分析

为了判断 RISC 过程中电荷转移类型及跃迁种类，我们对研究分子的单、三重激发态的跃迁轨道进行了分析。在 TD B3LYP/6-31+G（d，p）水平下优化了单、三重激发态几何。在此基础上，分析所研究分子处于激发单重态 $S_1$ 和三重态 $T_1$ 时涉及的前线分子轨道情况如图 5 所示：

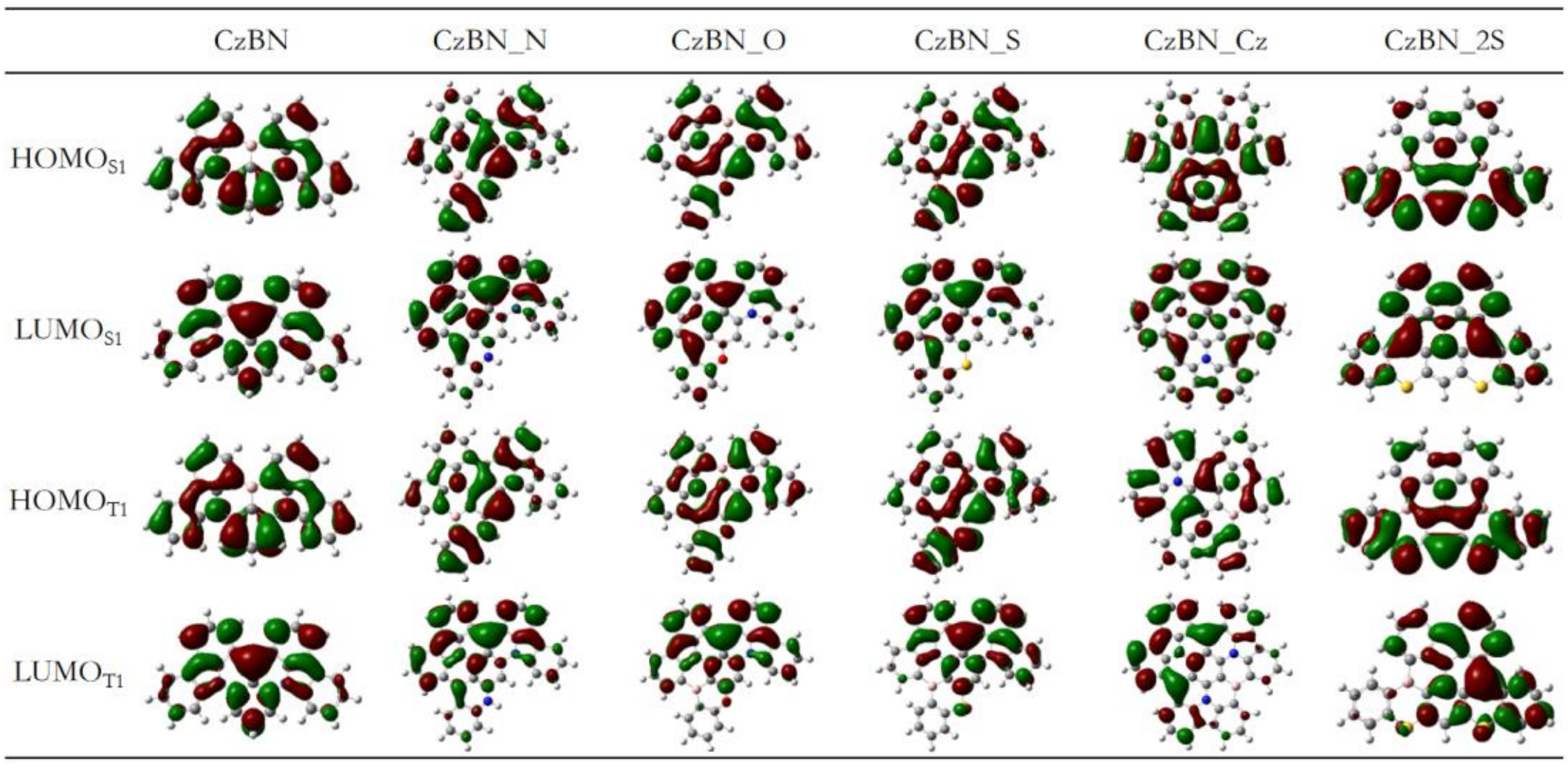

图 5　分子激发态 $S_1$ 和 $T_1$ 的跃迁分子轨道

为了分析方便，对分子结构的主要区域进行了划分（见图 6），计算了不同区域的 HOMO-LUMO 轨道占比，并分析其电荷转移类型。 结果列于表 9 至表中：

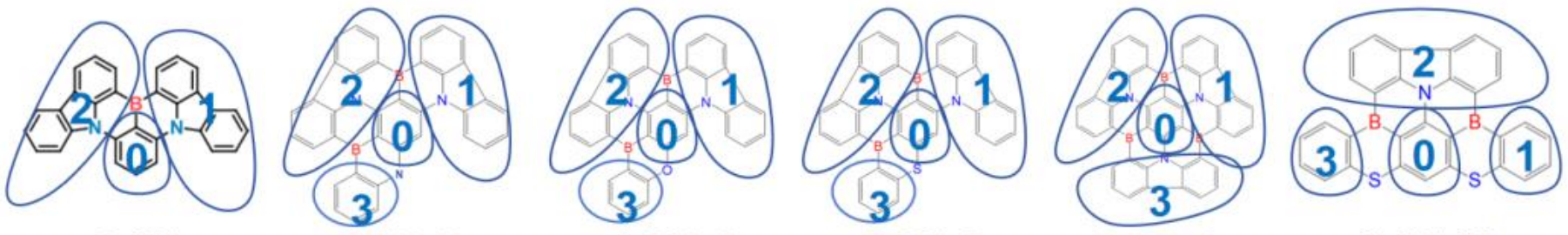

CzBN　CzBN_N　CzBN_O　CzBN_S　CzBN_Cz　CzBN_2S

图 6 研究分子的分子片划分图

表 8　TD(B3LYP/6-31+G（d，p）计算下的 $T_1$ 态电子跃迁

| | Excited states | $\lambda_{em}$ (nm) | $f$ | Main configurations |
|---|---|---|---|---|
| CzBN | $T_1$ | 536.45 | 0 | H-1→L(2%)；H→L(96%) |
| CzBN_N | $T_1$ | 513.43 | 0 | H-2→L(4%)；H→L(95%) |
| CzBN_O | $T_1$ | 551.62 | 0 | H-1→L+2(9%); H→L(91%) |
| CzBN_S | $T_1$ | 546.05 | 0 | H-1→L+2(8%); H→L(88%) |
| CzBN_Cz | $T_1$ | 581.35 | 0 | H→L (93%)；H→L+1(3%)<br>H→L+3(3%) |
| CzBN_2S | $T_1$ | 496.89 | 0 | H→L (99%) |

表 9　CzBN 各分子片在 HOMO 和 LUMO 中的占比

| CzBN | 0 | 1 | 2 | CT among the fragments |
|---|---|---|---|---|
| $HOMO_{S1}$ | 34.4 % | 32.8 % | 32.8 % | （0,1,2）→（0,1,2） |
| $LUMO_{S1}$ | 36.0 % | 32.0 % | 32.0 % | |
| $HOMO_{T1}$ | 32.2 % | 33.9 % | 33.9 % | （0,1,2）→（0,1,2） |
| $LUMO_{T1}$ | 36.4 % | 31.8 % | 31.8 % | |

表 10　CzBN_N 各区域 HOMO-LUMO 轨道占比

| CzBN_N | 0 | 1 | 2 | 3 | CT among the fragments |
|---|---|---|---|---|---|
| $HOMO_{S1}$ | 32.0 % | 25.0 % | 21.0 % | 22.0 % | （0,1,2,3）→（1,2） |
| $LUMO_{S1}$ | 8.0 % | 43.4 % | 44.6 % | 4.0 % | |
| $HOMO_{T1}$ | 31.2% | 26.4 % | 21.4 % | 21.0 % | （0,1,2,3）→（1,2） |
| $LUMO_{T1}$ | 8.3% | 46.7 % | 41.5 % | 3.5 % | |

表 11　CzBN_O 各区域 HOMO-LUMO 轨道占比

| CzBN_O | 0 | 1 | 2 | 3 | CT among the fragments |
|---|---|---|---|---|---|
| $HOMO_{S1}$ | 31.9 % | 28.9 % | 30.4 % | 8.8 % | （0,1,2）→（1,2） |
| $LUMO_{S1}$ | 7.8 % | 38.5 % | 48.1 % | 5.6 % | |
| $HOMO_{T1}$ | 27.2 % | 29.8 % | 31.1 % | 11.9 % | （0,1,2,3）→（1,2） |
| $LUMO_{T1}$ | 7.7 % | 43.6 % | 47.5 % | 1.2 % | |

表 12　CzBN_S 各区域 HOMO-LUMO 轨道占比

| CzBN_S | 0 | 1 | 2 | 3 | CT among the fragments |
|---|---|---|---|---|---|
| $HOMO_{S1}$ | 29.9 % | 31.9 % | 30.9 % | 7.3 % | （0,1,2）→（1,2） |
| $LUMO_{S1}$ | 8.7 % | 42.0 % | 44.9 % | 4.4% | |
| $HOMO_{T1}$ | 26.6 % | 32.8 % | 27.2 % | 13.4 % | （0,1,2,3）→（1,2） |
| $LUMO_{T1}$ | 7.5 % | 49.4 % | 42.3 % | 0.8 % | |

表 13　CzBN_Cz 各区域 HOMO-LUMO 轨道占比

| CzBN_Cz | 0 | 1 | 2 | 3 | CT among the fragments |
|---|---|---|---|---|---|
| $HOMO_{S1}$ | 25.0 % | 21.5 % | 21.0 % | 32.5 % | （0,1,2,3）→（1,2,3） |
| $LUMO_{S1}$ | 3.7 % | 41.0 % | 35.7 % | 19.6 % | |
| $HOMO_{T1}$ | 17.6 % | 23.2 % | 32.0 % | 27.2 % | （0,1,2,3）→（1,2,3） |
| $LUMO_{T1}$ | 5.5 % | 20.6 % | 52.6 % | 21.3 % | |

表 14　CzBN_2S 各区域 HOMO-LUMO 轨道占比

| CzBN_2S | 0 | 1 | 2 | 3 | CT among the fragments |
|---|---|---|---|---|---|
| $HOMO_{S1}$ | 7.3 % | 19.8 % | 52.0 % | 20.9 % | （1,2,3）→（0,1,2,3） |
| $LUMO_{S1}$ | 31.3 % | 14.0 % | 31.2 % | 23.5 % | |
| $HOMO_{T1}$ | 13.3 % | 43.7 % | 41.0 % | 2.0 % | （0,1,2）→（0,2,3） |
| $LUMO_{T1}$ | 27.1 % | 6.5 % | 47.3 % | 19.1 % | |

众所周知，对于 MR-TADF 材料分子，有效的 HOMO-LUMO 轨道分离可以促进单三态能隙的降低，进而有益于 RISC 速率的提升。若 RISC 过程通过 $^3$SRCT→$^1$LRCT 跃迁完成，该过程需要跨越的单三态能垒小，因此表现为大的 SOC，以及大的 RISC 速率。相反，若 RISC 过程通过 $^3$SRCT→$^1$SRCT 跃迁完成，该过程需要克服三重态和单重态之间固有的禁阻，因此表现为极小的 SOC，以及极小的 RISC 速率。本研究通过电荷跃迁分析揭示了分子结构修饰对 RISC 过程的调控机制：如表 9 所示，CzBN 分子 $S_1$ 和 $T_1$ 态都是由离域在分子整体的 HOMO 到 LUMO 跃迁组成（0,1,2 到 0,1,2 片段），因此 CzBN 分子呈现典型的 $^3$SRCT→$^1$SRCT 跃迁特征，这种短程电荷转移（CT），轨道分离度小，导致单三重激发态能隙大，SOC 值较低和 RISC 速率受限。经结构修饰后，新设计分子表现出显著的电荷跃迁特性转变—CzBN_N 实现 $^3$LRCT（0,1,2,3 到 1,2 片段）→$^1$LRCT（0,1,2,3 到 1,2 片段）跃迁，CzBN_O/S 形成 $^3$LRCT（0,1,2,3 到 1,2 片段）→$^1$SRCT（0,1,2,3 到 1,2 片段）跃迁，CzBN_2S 形成 $^3$LRCT（0,1,2,到 0,2,3 片段）→$^1$SRCT（1,2,3 到 0,2,3 片段），CzBN_Cz 则保持 $^3$SRCT→$^1$SRCT 主通道但引入少量 LRCT 成分（当始末态轨道区域差值为单一区域时，该电荷跃迁更倾向于 SRCT；而当轨道跨越两个或更多区域时，则更倾向于 LRCT）。这种电子结构重构产生三重优化效应：（1）SOC 矩阵元增强；（2）有效 $\Delta E_{ST}$ 减小；（3）构建多重 RISC 通道（$^3$LRCT→$^1$SRCT/$^1$LRCT）。分析表明，新设计分子的 RISC 速率将得到显著提升。

## 3.6　速率计算

理论计算了分子的单重激发态能量、三重激发态能量及单三重态能量差，SOC 以及重组能，最后通过爱因斯坦公式计算了荧光辐射跃迁速率（$k_r$），通过马库斯公式计算了系间窜越速率和反系间窜越速率（$k_{ISC}$ 和 $k_{RISC}$）如下

表：

表 15　所研究分子能级，重组能，SOC 以及速率计算

| | CzBN | CzBN_N | CzBN_O | CzBN_S | CzBN_Cz | CzBN_2S |
|---|---|---|---|---|---|---|
| $S_1$ (eV) | 3.06 | 3.21 | 3.29 | 3.15 | 3.32 | 3.11 |
| $T_1$ (eV) | 2.94 | 3.11 | 3.19 | 3.08 | 3.22 | 2.99 |
| $\Delta E_{S1T1}$ (eV) | 0.12 | 0.10 | 0.10 | 0.07 | 0.10 | 0.12 |
| $SOC_{S1}$ ($cm^{-1}$) | 0.17 | 0.13 | 0.25 | 0.60 | 0.17 | 0.16 |
| $SOC_{T1}$ ($cm^{-1}$) | 0.10 | 0.13 | 0.23 | 0.60 | 0.11 | 0.57 |
| $\lambda_{S1}$ ($cm^{-1}$) | 220 | 224 | 2337 | 2009 | 2071 | 945 |
| $\lambda_{T1}$ ($cm^{-1}$) | 342 | 342 | 1852 | 1149 | 2229 | 625 |
| $k_r$ ($s^{-1}$) | $1.31\times10^8$ | $8.97\times10^7$ | $1.40\times10^8$ | $1.24\times10^8$ | $2.32\times10^8$ | $1.18\times10^8$ |
| $k_{ISC}$ ($s^{-1}$) | $1.09\times10^7$ | $1.05\times10^7$ | $1.70\times10^7$ | $1.72\times10^8$ | $4.28\times10^6$ | $2.11\times10^7$ |
| $k_{RISC}$ ($s^{-1}$) | $1.14\times10^4$ | $8.98\times10^4$ | $1.52\times10^5$ | $3.48\times10^6$ | $6.83\times10^4$ | $2.76\times10^6$ |

由表 15 可知，设计分子 CzBN_N/O/S/Cz 的 $\Delta E_{ST}$ 分别降低至 0.10/0.10/0.07/0.10 eV（CzBN：0.12 eV），这是由于结构修饰显著改变了分子激发态特性，同时也源于多重杂原子杂环介导的 π 共轭扩展导致的 HOMO-LUMO 离域程度增大，以及轨道重叠度降低，而 CzBN_2S 因结构对称性与 CzBN 相似，其 $\Delta E_{ST}$ 保持 0.12 eV。所有新设计分子的 $S_1/T_1$ 态能量及 SOC 均因原子数和核电荷数增加而升高，其中 CzBN_O/S/2S 的 SOC 提升 2-3 倍，而 CzBN_N/Cz 基本不变。重组能分析显示 CzBN_O/S/Cz/2S 增加，但 CzBN_N 基本不变。最重要的是，新设计分子实现了 RISC 速率显著提升，特别是 CzBN_S、CzBN_2S 达 $3.48/2.76\times10^6$ $s^{-1}$（较 CzBN 提高 100 倍），这归因于 $\Delta E_{ST}$ 减小与 SOC 增强（0.57/0.60 $cm^{-1}$）的协同效应，突破了传统 MR-TADF 材料 $k_{RISC}$ 小的瓶颈。

## 4　结论

本研究通过 B-N-B 骨架的 π 共轭扩展及 B-N 杂环、B-O/S 杂环、B-N-B 杂环及双 S 杂环融合修饰，理论设计出 CzBN_N/O/S/Cz/2S 系列 MR-TADF 分子。理论计算证实：（1）设计分子保持着刚性平面 π 共轭结构和 MR 特性；（2）杂环融合结构修饰 CzBN 使 $\Delta E_{ST}$ 显著降低（CzBN_S 达 0.07 eV），SOC 提升 2-3 倍；）（3）RISC 速率突破性提升（CzBN_S 达 $3.48\times10^6$ $s^{-1}$，提升 100 倍）；

（4）维持窄带发射特性（FWHM≈40 nm）并实现蓝移（400-450 nm）; (5) Stocks shift 小（低于 25nm），激子利用率高。该设计策略通过协同调控 $\Delta E_{ST}$ 减小和 SOC 增强，提高了 RISC 窜越速率 $k_{RISC}$, 有效解决了三线态激子淬灭问题，同时减小了 Stokes shift（对于 CzBN_O/S/Cz， Stokes shift 都小于 21nm），提升了激子利用率。本研究建立的计算方法体系为 MR-TADF 材料的理性设计提供了可靠的理论依据，所提出的分子设计策略为高性能 OLED 器件的开发奠定了重要基础。

## 5 参考文献

[1] Ahn D H, Kim S W, Lee H, et al. Highly efficient blue thermally activated delayed fluorescence emitters based on symmetrical and rigid oxygen-bridged boron acceptors[J]. Nature Photonics, 2019, 13(8): 540-546.

[2] Luo X F, Qu Z Z, Han H B, et al. Carbazole‐Based Iridium (III) Complexes for Electrophosphorescence with EQE of 32.2% and Low Efficiency Roll‐Off[J]. Advanced Optical Materials, 2021, 9(3): 2001390.

[3] Active tuning of surface phonon polariton resonances via carrier photoinjection

[4] Yurash B, Nakanotani H, Olivier Y, et al. Photoluminescence quenching probes spin conversion and exciton dynamics in thermally activated delayed fluorescence materials[J]. Advanced Materials, 2019, 31(21): 1804490.

[5] Song F, Xu Z, Zhang Q, et al. Highly efficient circularly polarized electroluminescence from aggregation‐induced emission luminogens with amplified chirality and delayed fluorescence[J]. Advanced Functional Materials, 2018, 28(17): 1800051.

[6] Santoro F, Lami A, Improta R, et al. Effective method for the computation of optical spectra of large molecules at finite temperature including the Duschinsky and Herzberg–Teller effect: The Qx band of porphyrin as a case study[J]. The Journal of chemical physics, 2008, 128(22).

[7] Poitras D, Kuo C C, Py C. Design of high-contrast OLEDs with microcavity effect[J]. Optics Express, 2008, 16(11): 8003-8015.

[8] Hatakeyama T, Shiren K, Nakajima K, et al. Ultrapure Blue Thermally Activated Delayed Fluorescence Molecules: Efficient HOMO-LUMO Separation by the Multiple Resonance Effect[J]. Advanced Materials (Deerfield Beach, Fla.), 2016, 28(14): 2777-2781.

[9] Yang M, Park I S, Yasuda T. Full-color, narrowband, and high-efficiency electroluminescence from boron and carbazole embedded polycyclic heteroaromatics[J]. Journal of the American Chemical Society, 2020, 142(46): 19468-19472.

[10] Jiang P, Miao J, Cao X, et al. Quenching‐resistant multiresonance TADF emitter realizes 40% external quantum efficiency in narrowband electroluminescence at high doping level[J]. Advanced Materials, 2022, 34(3): 2106954.

[11] Zhang Y, Li G, Wang L, et al. Fusion of multi - resonance fragment with conventional polycyclic aromatic hydrocarbon for nearly BT. 2020 green emission[J]. Angewandte Chemie International Edition, 2022, 61(24): e202202380.
[12] Van Donkelaar A, Martin R V, Brauer M, et al. Global estimates of fine particulate matter using a combined geophysical-statistical method with information from satellites, models, and monitors[J]. Environmental science & technology, 2016, 50(7): 3762-3772.
[13] Hu S, Tian Y, Lin Y, et al. High-efficiency and long-lifetime deep-blue organic light-emitting diode with a maximum external quantum efficiency of 20.6% and CIEy of 0.04[J]. Dyes and Pigments, 2022, 205: 110548.
[14] Cao X, Pan K, Miao J, et al. Manipulating exciton dynamics toward simultaneous high-efficiency narrowband electroluminescence and photon upconversion by a selenium-incorporated multiresonance delayed fluorescence emitter[J]. Journal of the American Chemical Society, 2022, 144(50): 22976-22984.
[15] Hu Y X, Miao J, Hua T, et al. Efficient selenium-integrated TADF OLEDs with reduced roll-off[J]. Nature Photonics, 2022, 16(11): 803-810.
[16] Pratik S M, Coropceanu V, Brédas J L. Enhancement of thermally activated delayed fluorescence (TADF) in multi-resonant emitters via control of chalcogen atom embedding[J]. Chemistry of Materials, 2022, 34(17): 8022-8030.
[17] Huang Z, Xie H, Miao J, et al. Charge transfer excited state promoted multiple resonance delayed fluorescence emitter for high-performance narrowband electroluminescence[J]. Journal of the American Chemical Society, 2023, 145(23): 12550-12560.
[18] Yang M, Shikita S, Min H, et al. Wide - range color tuning of narrowband emission in multi - resonance organoboron delayed fluorescence materials through Rational Imine/Amine Functionalization[J]. Angewandte Chemie International Edition, 2021, 60(43): 23142-23147.
[19] Situ Z, Li X, Gao H, et al. Accelerating Intersystem Crossing in Multiresonance Thermally Activated Delayed Fluorescence Emitters via Long-Range Charge Transfer[J]. The Journal of Physical Chemistry Letters, 2024, 15(15): 4197-4205.
[20] Frisch M J, Trucks G W, Schlegel H B, et al. Gaussian 16 Rev. C.01. Wallingford, CT, 2016.
[21] Kállay, Mihály, et al. "The MRCC program system: Accurate quantum chemistry from water to proteins." The Journal of chemical physics 152.7 (2020).
[22] Hatarik, Robert, et al. "Analysis of the neutron time-of-flight spectra from inertial confinement fusion experiments." Journal of Applied Physics 118.18 (2015).
[23] Neese F. The ORCA program system[J]. Wiley Interdisciplinary Reviews: Computational Molecular Science, 2012, 2(1): 73-78.
[24] Zhang J, Yuan H, Feng S, et al. Theoretical studies on electroluminescent mechanism of a series of thermally activated delayed fluorescence emitters possessing asymmetric-triazine-cored triads[J]. Spectrochimica Acta Part A: Molecular and Biomolecular Spectroscopy, 2018, 202: 102-106.
[25] Gao Y-J, Chen W-K, Zhang T-T, et al. Theoretical Studies on Excited-State

Properties of Au(III) Emitters with Thermally Activated Delayed Fluorescence[J/OL]. The Journal of Physical Chemistry C, 2018, 122(48): 27608-27619.